\input{aipcheck}

\documentclass[
    sort&compress
    ,final            % use final for the camera ready runs
  ]
  {aipproc}

\layoutstyle{6x9}

\begin{document}

\title{Digging for formational clues in the halos of early-type galaxies}


\author{Aaron. J. Romanowsky}{
  address={UCO/Lick Observatory, University of California, Santa Cruz, CA 95064, USA}
}

\begin{abstract}
Many of the fundamental properties of early-type galaxies (ellipticals and lenticulars)
can only be accessed by venturing beyond their
oft-studied centers into their large-radius halo regions.
Advances in observations of kinematical tracers allow early-type halos
to be increasingly well probed.
This review focuses on recent findings on angular momentum and dark matter content,
and discusses some possible implications for galaxy structure and formation.
\end{abstract}

\maketitle

%%%%%%%%%%%%%%%%%%%%%%%%%%%%%%%%%%%%%%%%%%%%
%% MAINMATTER
%%%%%%%%%%%%%%%%%%%%%%%%%%%%%%%%%%%%%%%%%%%%

\section{Introduction}

The Milky Way's halo has for a century been a Rosetta stone for
deciphering the structure and formational history of our home galaxy
by using multitudes of individual stars as chemo-dynamical tracers.
Recent observational advances 
have extended this ``archaeological'' approach to Local Group galaxies including the massive spiral M31.
%AJR: cite?
The ultimate goal is the study of resolved stellar populations in
all galaxy types and environments,
whose fruition will require the next generation of giant telescopes.
In the meantime a great deal of information can be learned through the alternatives
of integrated stellar light 
\citep{2009MNRAS.398...91P,2009MNRAS.398..561W,2009AJ....138.1417T,2009MNRAS.400.2135F}, 
and resolved tracers such as planetary nebulae (PNe; 
\citep{2009MNRAS.393..329N,2009MNRAS.394.1249C,2009MNRAS.395...76D}) and globular clusters (GCs; \citep{2009AJ....137.4956R}).
%AJR: Merrett ref?
%AJR: more GCs refs? ARAA?  Beasley?

These observational tools are particularly apt for delving into
massive early-type galaxies (ellipticals and lenticulars),
whose evolutionary histories are challenging to understand
in a cosmological context.
%which are an ongoing focus of puzzlement.
%which will be the focus of the discussion from here on.
A simple paradigm where gas-rich disk galaxies fade to lenticulars,
or merge into ellipticals, is giving way to a more complex picture
where minor mergers and secular processes play critical roles
\citep{2008ApJ...688...67E,2009ApJ...703..785D,2009MNRAS.396.1972P,Hopkins10}.
%Those spheroid-dominated systems are classically considered to be the
%descendents of major mergers between gas-rich disk galaxies, but recent 
To decrypt the varied genealogies of low redshift galaxies,
some of the key fossil clues 
%Some of the key properties that can be probed
are the halo distributions of angular momentum
and mass (including dark matter), orbit structures, metallicity gradients,
and substructures such as streams and shells.

The cornerstones of such halo studies are high-quality, deep, wide-field imaging and
spectroscopy, detailed modeling, and comparison to simulations of galaxy formation.
Large surveys of early-type galaxy halos are currently underway at the Keck, Subaru,
and William Herschel telescopes, using stars, GCs, and PNe:  
SMEAGOL, SLUGGS, and the PN.S Elliptical Galaxy Survey
\citep{2009MNRAS.398...91P,2009MNRAS.400.2135F,2009AJ....137.4956R,2009MNRAS.393..329N,2009MNRAS.394.1249C,2009MNRAS.395...76D}.
These projects complement the central surveys by SAURON \citep{2007MNRAS.379..401E,2007MNRAS.379..418C}
and enable the construction of comprehensive global galactic models.

\section{Angular Momentum}

The central regions of early- and late-type galaxies differ dramatically in their
rotational properties, which may reflect differences in angular momentum conservation
during their assembly histories \citep{1983IAUS..100..391F}.
Among the early-types, there are two broad sub-types:
the fainter, disky, fast-rotators with cuspy centers; and the
brighter, boxy, slow-rotators with central cores.
The SAURON survey has dramatically demonstrated this distinction based on
kinematics and dynamics, motivating 
an angular momentum metric
%a specific angular momentum parameter $\lambda_R$
as the primary classifier for galaxies
\citep{2007MNRAS.379..401E,2007MNRAS.379..418C}.
%AJR: KB96? 

The fast-rotators are characterized as oblate axisymmetric systems
which are likely shaped by dissipative processes, as in a major gas-rich (``wet'') merger.
The slow-rotators are triaxial with surprisingly isotropic orbits, with so far no
formation model that fully explains their properties, although it is plausible that
they originated in multiple mergers at high redshift \citep{2008ApJ...685..897B}.

The advent of larger-radius kinematical data now suggests that
standard rotation-based classifications may be relevant only for the central regions.
The rotation profiles outside of an effective radius ($R_{\rm eff}$) are diverse, including common
cases of ``fast-rotators'' where the outer rotation amplitude drops dramatically (Fig.~1),
yielding global specific angular momentum values that are comparable to those of the slow
rotators (and still much smaller than in late-types).
There are also hints of kinematic twists appearing at large radii \citep{2009MNRAS.394.1249C},
suggesting an onset of triaxiality.

\begin{figure}
  \includegraphics[height=.33\textheight]{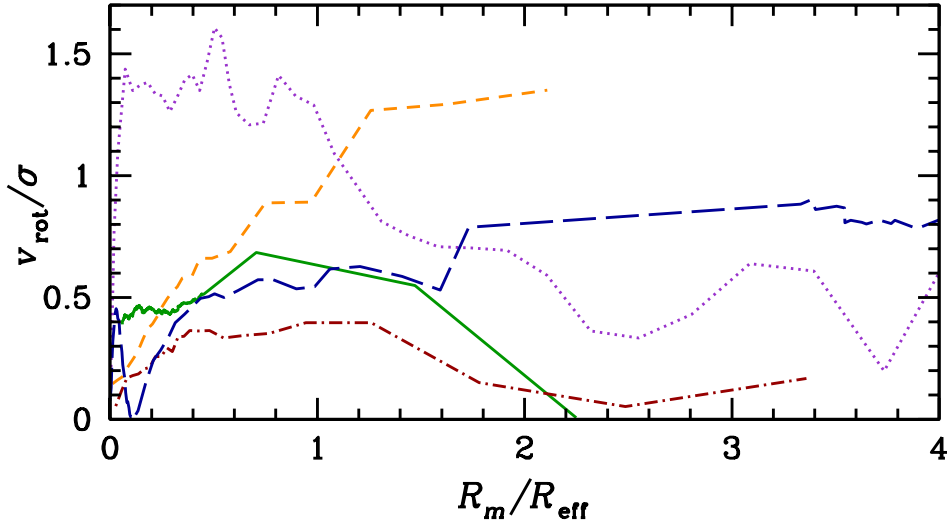}
  \caption{Rotation dominance parameter (major-axis rotation amplitude over velocity
dispersion) versus radius 
%(in units of the effective radius)
for a sample of early-type galaxies, 
%chosen to show
selected to illustrate
the range of outer rotation profiles observed \citep{2009MNRAS.394.1249C,2009MNRAS.398...91P,2009MNRAS.398..561W}.
%AJR: reference Jacob here?
All cases are classified as fast-rotators
based on their central regions.}
\end{figure}

These observations concord with photometric results for many disky
ellipticals to transition to rounder and boxier spheroids at large radii.
There is a long-standing model wherein all early-types can be
characterized primarily as extended bulges,
often with an additional central disklike component
\citep{1990ApJ...362...52R,1995A&A...293...20S,2008MNRAS.390...93K}.
%\citep{1990ApJ...362...52R,1994MNRAS.270..325C,1995A&A...293...20S,2008MNRAS.390...93K}.
The large-radius kinematics dramatically confirm this picture
and suggest that the bulge rotation tends to be fairly {\it slow} 
outside the central regions (where the disk and bulge may be coupled).

%%%%%%%%%%%%%%%%%%%%%%%%%%%%%%%%%%%%%%%%%%%%
%% Sample figure:
%%
%% The option [height=...] scales the picture to the given height,
%% without it it would be printed at its nominal size
%%%%%%%%%%%%%%%%%%%%%%%%%%%%%%%%%%%%%%%%%%%%

%The weakly-coupled two-component nature of early-type galaxies should provide
%constraints on their formation scenarios, once the observational details are
%better understood.
A number of formational possibilities are suggested by this weakly-coupled
two-component picture of early-type galaxies.
Recent analysis of 1:1 wet merger simulations demonstrates that in this
classic scenario, an observable decoupling between the central and outer
regions is naturally expected---reflecting the wet and dry components of the
merger, respectively \citep{Hoffman10}.
Growth of the outer envelope by minor mergers may produce a similar effect.
Another possibility is that stream-fed high-redshift ``wild disks''
\citep{2008ApJ...688...67E,2009ApJ...703..785D}
might build up a bulge with rotation decreasing outwards.
%where dissipation creates increasing rotation towards the center.
It remains to be seen in detail how these various scenarios' 
predictions for rotation amplitudes and twists square with observations.
One potential degeneracy breaker is the use of GC subpopulations, since these
would originate in different components of the galactic progenitors
\citep{2010MNRAS.401L..58B,Hoffman10}.
%AJR: reference Jacob here?

\section{Dark matter}

%Halos of dark matter (DM) are thought to be the bedrock of galaxy formation, but the 
The dark matter (DM) content of ordinary early-type galaxies is much more poorly known than for late-types.
Although some constraints have been provided by gravitational lensing and X-ray gas emission,
there is a critical need for detailed DM profiles in an unbiased sample of galaxies.
The radially extended dynamics of stars, PNe, and GCs are starting to fill this void.
Despite systematic modeling difficulties in deriving robust DM constraints,
independent efforts are so far returning consistent results
\citep{2009MNRAS.395...76D,2009MNRAS.398..561W}, 
and further progress will come through combining multiple tracers in the same galaxies.

The hodgepodge of results so far available in the literature paint a startling picture
of the DM content in early-types (Fig.~2).
The slow-rotators appear to have much higher halo concentrations than the fast-rotators,
with the theoretical prediction treated as a zone of avoidance.
Other inferences from more central regions are not consistently supportive of these large-radius
results \citep{2006MNRAS.366.1126C,2009ApJ...691..770T,2009MNRAS.396.1132T}, 
%AJR: Barnabe 2009; Chen & McGaugh 2008
which may in fact be highly skewed by observational selection effects---demonstrating
the need for an unbiased survey.

\begin{figure}
  \includegraphics[height=.31\textheight]{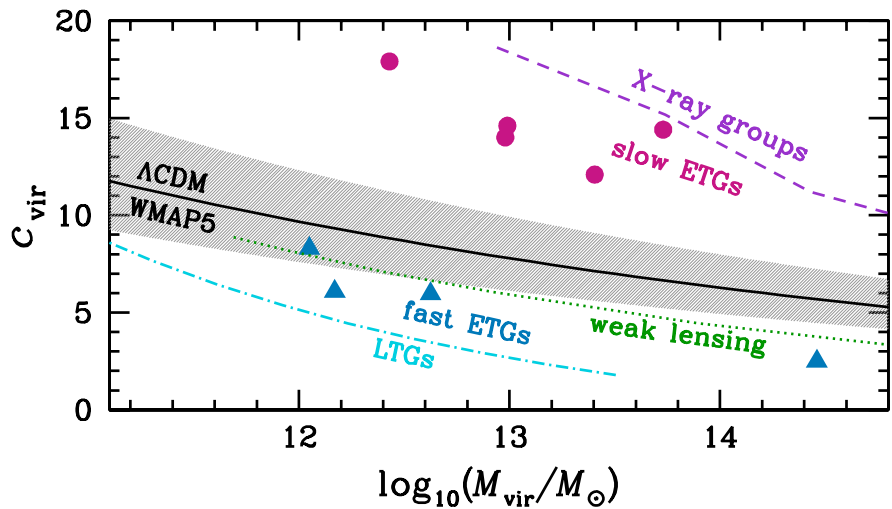}
  \caption{Dark matter halo masses and concentrations for early-type galaxies based on dynamics
\citep{2009MNRAS.393..329N}, with shading showing updated theoretical predictions
%mean theoretical predictions shown as a curve with the 1-$\sigma$ scatter 
\citep{2008MNRAS.391.1940M}.  
The slow- and fast-rotators are shown with different symbols and labeled
accordingly.  The error bars in this parameter space are large and correlated
because of the extrapolations to the virial radius, but
a re-casting to well-constrained central dark matter parameters
shows a similar dichotomy between the fast- and slow-rotators.
For comparison, independent results are shown for X-ray groups (which typically have central
slow-rotators) and for late-type galaxies.  A mean trend for both early- and late-types based
on weak gravitational lensing is also included.}
\end{figure}

If the dichotomy does hold up in an unbiased sample, it would not be explained merely
by some kind of preferential population of DM halos.
Instead, systematic differences would be implied in the interplay between baryons and
DM during galaxy assembly, affecting the central DM densities.
Contraction of the halo during baryonic collapse would have been efficient in
slow rotators, but not in fast-rotators or in late-types (see Fig.~2).
There have been many mechanisms proposed for inefficient halo contraction, with
the dynamical effects of lumpy accretion emerging as a major contender \citep{2009ApJ...697L..38J}.
In this case, a history of smoother accretion might be implied for the slow rotators.

Another puzzle involves the halo {\it orbits} of stars and GCs:
fast rotators show radial bias as expected 
\citep{2009MNRAS.393..329N,2009MNRAS.395...76D,2009MNRAS.398..561W},
but the slow rotators may be isotropic or {\it tangentially}-biased
\citep{2009AJ....137.4956R}.
Piecing together all these clues should help decipher galaxies' formational pathways.
%In the case, the puzzle is how
%The question then becomes, why would these effects not operate for the slow rotators?
%of late-type galaxies \citep{2007ApJ...654...27D}, and
%Late-type galaxies are currently thought to have somewhat
%low central DM densities suggesting dynamical feedback processes which prevented
%DM halo contraction from baryonic collapse \citep{2007ApJ...654...27D}.  
%Similar results for fast-rotator early-types (Fig.~2) suggest formational histories
%that are related.

%%%%%%%%%%%%%%%%%%%%%%%%%%%%%%%%%%%%%%%%%%%%%%%%
%% BACKMATTER
%%%%%%%%%%%%%%%%%%%%%%%%%%%%%%%%%%%%%%%%%%%%%%%%

\begin{theacknowledgments}
I wish to thank my collaborators for the many efforts and ideas contributing
to the work discussed here.
Support provided by NSF Grants AST-0808099 and AST-0909237.
\end{theacknowledgments}




%\bibliography{aromanowsky}

%  \typeout{******************************************}
%  \typeout{** Please run "bibtex \jobname" to optain}
%  \typeout{** the bibliography and then re-run LaTeX}
%  \typeout{** twice to fix the references!}
%  \typeout{******************************************}

\end{document}